**Brain-inspired, interpretable, resonant recurrent neural networks**

*Mark A. Kramer*

*Department of Mathematics and Statistics & Center for Systems Neuroscience*
*Boston University, Boston, MA, 02215*

*Department of Applied Mathematics and Statistics*
*Johns Hopkins University, Baltimore, MD, 21218*

## Abstract

Traditional artificial neural networks consist of nodes with non-oscillatory dynamics. Biological neural networks, on the other hand, consist of oscillatory components embedded in an oscillatory environment. Motivated by this feature of biological neurons, we describe a neural network framework with explicit damped, oscillatory node dynamics. We express the oscillatory dynamics using two history dependent terms to connect these dynamics with standard recurrent neural network formulations, apply physical constraints from observed brain dynamics to choose the oscillatory frequencies, and stationary constraints to reduce the number of free parameters. We then optimize and illustrate network performance by classifying hand-written digits and simulated neuronal spike train activity and show that these oscillatory network elements support accurate classification with few trainable parameters. Choosing oscillator frequencies according to a proposed theory for brain rhythms improves classification accuracy compared to alternative frequency configurations and compared to standard recurrent neural network frameworks with comparable numbers of parameters. Compared to existing approaches, the proposed *resonant* recurrent network (RRN) utilizes oscillatory dynamics expressed as a straightforward extension of standard recurrent neural networks, produces interpretable features for classification, and performs well with few parameters when oscillator frequencies follow a configuration observed *in vivo*. We propose that RRNs may serve as efficient, biologically inspired building blocks to achieve complex goals in biological and artificial neural networks.

## Introduction

Biological systems rely on stable dynamics, which typically oscillate when perturbed as compensatory mechanisms prevent divergent activity. For example, a tree perturbed by the wind or blood glucose perturbed by sugar oscillate in their return to equilibrium. These intrinsically oscillatory systems exist in the Earth's environment, which imposes oscillations across temporal scales, from ocean waves, to diurnal cycles, to lunar cycles, to yearly cycles. Biological systems on Earth communicate with oscillations (e.g., acoustic disturbances like speech, bird-calls, or dolphin clicks) and interact with the world to produce and control oscillations (e.g., breathing [1], movement [2], and digestion [3]). Importantly, the building blocks of animal intelligence – neurons – often exhibit oscillatory dynamics alone and in circuits with other neurons [4] to support functions like memory consolidation [5], executive control [6], and navigation [7].

Features of the brain's biological processes have motivated the development of artificial neural networks. Like biological neurons, artificial neurons respond when sufficient input drives stereotyped output, which subsequently excites or inhibits other neurons and impacts the connections between neurons [8], [9], [10]. To process spatially structured input images, biological neurons in early visual cortex respond preferentially to localized regions and features of the visual field [11], which motivated the development of local receptive fields in convolutional neural networks [12]. To process temporally structured input signals, biological brains may exploit the temporal characteristics of individual neurons or neural population dynamics (e.g., rhythms [4], [13], resonance [14], [15], coherence [16], [17], and cross-frequency coupling [18], [19]). More recently, biological brain rhythms have motivated the development of artificial neural networks with imposed [20] or intrinsic [21] rhythmic structure. These recent works have shown that artificial neural networks with rhythmic features support multiplexing of complex inputs [20], and enhance learning speed, noise tolerance, and parameter efficiency [22].

Motivated by the utility of brain rhythms shown *in vivo* and *in silico* [4], [16], [23], we propose a modified recurrent neural network with rhythmic structure. Traditional recurrent neural networks are commonly written either as discrete-time difference equations in which the hidden state at time index $t$ depends on the state at $t-1$ [24], or as first-order continuous-time dynamical systems [25], [26]; while such formulations can exhibit oscillatory trajectories, these formulations do not explicitly parameterize intrinsic, frequency-selective modes. To include rhythmic activity, neural networks with explicit band-pass filters have been proposed [27], [28], and recent efforts have implemented networks as complex valued equations [29] or second order differential equations modeling networks of oscillators [21], [22], [30]. While these oscillatory networks benefit from physical interpretability (e.g., as damped driven oscillators) and direct mathematical analysis via classical coupled-oscillator theory [31], these sophisticated formulations differ from traditional recurrent neural network expressions; a simplified framework may improve interpretability between, and insights from, different artificial neural network strategies.

To that end, we propose a resonant recurrent network (RRN) consisting of coupled, damped, driven oscillators that respond to time series input at different frequencies. Unlike previous approaches, we explicitly formulate the oscillator dynamics as a discrete-time recurrent neural network with two history dependent terms. In this way, we link the physical oscillator dynamics to traditional recurrent neural network and statistical formulations, and we apply biological and stability constraints to limit the number of hyperparameters. To illustrate RRN performance, we analyze data from a well-established task: classification accuracy of hand-written digits. We show that, consistent with alternative coupled oscillator network approaches [21], [22], [30], the RRN performs well in this classification tasks compared to existing approaches (a traditional recurrent neural network and long-short term memory [LSTM] network) with a comparable number of parameters. In addition, we show that choosing RRN oscillatory frequencies to mimic one proposed theory for the brain's rhythmic organization results in networks that outperform alternative frequency configurations. We find consistent results for classification of synthetic modulated or unmodulated spike train data. The proposed conceptual framework links techniques from statistics (spectral estimation theory) and physics (coupled oscillator dynamics) to a traditional recurrent neural network formulation and provides a simple, biologically-motivated computing strategy with few parameters and immediately interpretable dynamics.

## Results

### A resonant recurrent network with stability constraints

The RRN consists of $K$ coupled nodes. We begin by expressing the dynamics of the $k^{th}$ individual (i.e., uncoupled) node in the RRN as a damped, driven oscillator:

$$\ddot{x}^k = -2\,\beta_k\,\dot{x}^k - \omega_k^2\,x^k + \xi^k(t) + U^k(t)\ , \tag{1}$$

where $x^k$ is the position, $\beta_k$ is the damping constant, and $\omega_k$ is the natural frequency of the $k^{th}$ oscillator. We include an optional noise term $\xi^k(t)$ [27] to the $k^{th}$ oscillator evaluated at time $t$ to create spontaneous activity consistent with biological neural networks [13]. Finally, we include an external input $U^k(t)$ to the $k^{th}$ oscillator at time $t$.

Replacing the derivatives with discrete approximations, Equation (1) becomes

$$x_t^k = \alpha_1^k\,x_{t-1}^k + \alpha_2^k\,x_{t-2}^k + \epsilon_t^k + u_t^k\ , \tag{2}$$

where $\alpha_1^k$ and $\alpha_2^k$ are functions of the physical parameters $\beta_k$ and $\omega_k$, and $\epsilon_t^k$ (normally distributed with zero mean and standard deviation $\sigma$) and $u_t^k$ are the noise and driving terms from Equation (1), respectively, scaled by the time interval of the discrete approximation; see Appendix A of [32]. We note that, in this way, the damped driven oscillator is equivalent to an autoregressive model of order 2 (i.e., an AR(2)) driven by external input $u_t^k$.

The complete RRN model consists of $K$ coupled nodes with dynamics

$$x_t^k = \alpha_1^k\,x_{t-1}^k + \alpha_2^k\,x_{t-2}^k + \epsilon_t^k + u_t^k + \sum_{j\neq k} W^{kj}\,f\left(u_t^k\,x_{t-1}^j\right) \tag{3}$$

where we update Equation (2) to include interactions between nodes. The matrix $W$ indicates the connectivity,

where $W^{kj}$ represents the connectivity from node $j$ to node $k$, and we set $W^{kk} = 0$ (i.e., no self-connections). We couple node dynamics using the product of the external input to node $k$ ($u_t^k$) and the position of node $j$ with nonlinearity $f(x) = \tanh(x)$. With this choice of coupling, the external input directly drives node $k$ and modulates the impact of node $j$ on node $k$ (consistent with the phenomenon of gain modulation [33], [34]).

The RRN model in Equation (3) supports multiple interpretations. First, we motivated the RRN in Equation (3) using a physical system: a coupled network of damped, driven harmonic oscillators [32]. Expressed as a discrete difference, the dynamical equation for the physical system becomes more like typical recursive neural networks, but not quite. Typical recursive neural networks (e.g., simple recursive neural networks [24]) or networks with advanced gates (e.g., LSTM [35] or GRU [36] networks) include dependence only on the previous state (i.e., one history dependent term). Here, motivated by the physical model of a damped oscillator, we extend this original formulation to include two history dependent terms; the inclusion of the new term ($\alpha_2^k \, x_{t-2}^k$) allows intrinsic oscillatory dynamics at each node with features (i.e., natural frequency and damping) we control through parameters $\alpha_1^k$ and $\alpha_2^k$.

We leverage these alterative formulations in different ways to support simplicity, enhance interpretability, and apply biological and mathematical constraints. We choose the formulation in Equation (3) – rather than formulating the systems as coupled oscillators – for its simplicity and consistency with typical recursive neural networks with one history dependent term (i.e., with $\alpha_2^k = 0$). To establish damped oscillatory dynamics at center or resonant frequencies $f_0$, we choose

$$\begin{aligned} \alpha_1^k &= 2\, r \cos\left(2\, \pi \frac{f_0}{F_s}\right) \\ \alpha_2^k &= -r^2 \end{aligned} \tag{4}$$

where $r$ is the damping parameter and $F_s$ is the sampling frequency of our discrete time system; in what follows, we set $r = 0.99999$ for narrowband oscillations and light damping, and we fix $F_s = 1000$ Hz. To enforce stationarity, we require the roots of the polynomial

$$1 - \alpha_1^k \, z^{-1} - \alpha_2^k \, z^{-2} = 0 \tag{5}$$

lie inside of the unit circle in the complex plane [37]. We additionally constrain $\alpha_1^k > 0$ to avoid sign-alternating dynamics and restrict frequencies to $f_0 < \frac{F_s}{2}$ to prevent aliasing. We set $f_0 \geq 2$ Hz to focus on rhythms in the traditional frequency bands studied *in vivo* [38], [39]. In what follows, we choose the frequencies $f_0$ based on different proposals for the organization of mammalian brain rhythms *in vivo*.

As an initial illustration of the RRN dynamics, we simulate a network with 11 nodes, choosing frequencies spaced by factors of the golden ratio $f_0 \in \{2, 3.2, 5.2, 8.5, 13.7, 22.2, 35.9, 58.1, 93.9, 152.0, 245.9\}$ Hz, motivated by observed distributions of brain rhythms *in vivo* [32], [39], [40]. We set the input to 0 ($u_t^k = 0$ in Equation 3) and remove all connections between nodes ($W = 0$ in Equation 3). The dynamics of each node then consist of damped oscillations (i.e., fading memory) with narrow spectral peaks (example theoretical spectra [37] with these parameters shown in Figure 1A). The noise ($\epsilon_t^k$ in Equation 3 with $\sigma = 0.005$) results in noisy, rhythmic fluctuations in the activity of each oscillator (Figure 1B). This example illustrates the interpretability of the RRN network dynamics as damped driven oscillators.

In what follows, we train RRNs to classify input signals (i.e., both $u_t^k$ and $W$ are no longer restricted to 0) and show how performance depends on node frequencies. We simplify by setting the noise to zero ($\epsilon_t^k = 0$ for all $k, t$ in Equation 3) and applying the same input to all network nodes, so that $u_t^k$ becomes $u_t$ in Equation 3. We note that, in doing so, we fix the input weight to 1 for all nodes, i.e., the same input signal drives all network elements. In the examples that follow, we apply only minimal preprocessing to each input signal (e.g., simple scaling). We compute per-node features as the time-averaged squared amplitude and pass these features through batch normalization and a linear readout to produce features for classification. Intuitively, we expect nodes will respond to the frequency content of the input signals and learn cross-frequency interactions to support classification. We show that RRNs outperform comparable recurrent neural networks in two classification tasks when the RRNs implement a biologically motivated choice of node frequencies.

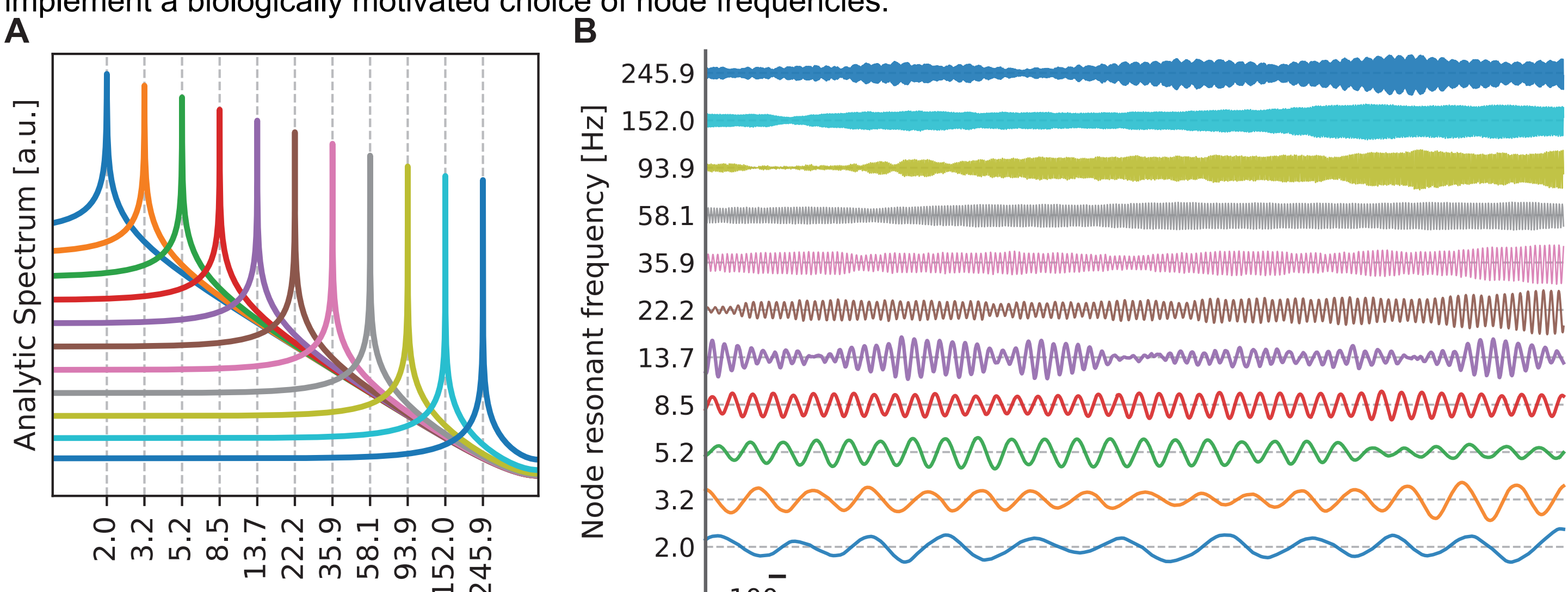

***Figure 1: The resonant recurrent network (RRN) consists of damped, coupled oscillators. (A)*** *Theoretical spectra computed for 11 nodes with center frequencies separated by factors of the golden ratio.* ***(B)*** *Example dynamics of this RRN driven by noise; colors correspond to the spectra in (A).*

**The resonant recurrent network – with golden ratio spacing between frequencies – outperforms alternative configurations and comparable architectures in a common classification task**

To characterize RRN performance, we analyzed the well-established problem of classifying the handwritten digits in the MNIST dataset (see *Methods: MNIST Dataset*). We started by selecting three configurations of node frequencies, motivated by observations from *in vivo* brain dynamics (Table 1). These configurations separate the brain's discrete rhythms by factors of Euler's number ($e \approx 2.718$) [41], two [42], or the golden ratio ($\phi \approx 1.618$) [32]. We trained and tested each configuration in the same way, with two hyperparameters (weight decay and label smoothing coefficient) selected for each architecture to maximize accuracy on held-out validation data from the training set (see *Methods: Hyperparameter selection*). Training estimated the connectivity between nodes ($W$ in Equation 3) and parameters for the linear classifier and batch normalization. We trained each architecture with 10 realizations of the MNIST data, using the same randomized train/validation data splits for each architecture, and computed the classification accuracy of the test data (not used in training or validation). We found that the RRN with frequencies separated by factors of the golden ratio - the *golden RRN* - performed with higher accuracy than the other two biologically motivated configurations (Table 1, $p < 1e^{-14}$ , Figure 2B). Moreover, the golden RRN outperformed a network with the same number of parameters but linearly spaced frequencies ($p = 4e^{-14}$, Table 1). Finally, compared to existing recurrent neural networks with a comparable number of parameters (a LSTM and standard Elman RNN), the golden RRN again performed with higher classification accuracy (LSTM: $p = 0.0098$; standard RNN: $p = 1.1e^{-11}$; see Table 1). Confusion matrices demonstrated comparable or superior performance of the golden RRN compared to the other architectures (per-class accuracy of the golden RRN exceeded or matched all other architectures except for digits 0 [LSTM and standard RNN] and 8 [standard RRN]; Figure 2C). We conclude that the RRN with frequencies separated by factors of the golden ratio outperformed other biologically motivated RRNs and existing recurrent neural network approaches with comparable numbers of parameters.

| | Node frequencies (Hz) | Parameters | Hyperparameters (decay, smoothing) | MNIST Accuracy Mean (SEM) | Spike train Accuracy Mean (SEM) |
|---|---|---|---|---|---|
| Golden (n=11) | $f_0 \in \{2, 3.2, 5.2, 8.5, 13.7, 22.2, 35.9, 58.1, 93.9, 152.0, 245.9\}$ | 252 | 0, 0 | 0.759 (0.007) | 0.635 (0.001) |
| Two (n=7) | $f_0 \in \{2, 4, 8, 16, 32, 62, 128\}$ | 136 | 0.01, 0 | 0.583 (0.007) | 0.622 (0.001) |
| Euler's (n=5) | $f_0 \in \{2, 5.4, 14.8, 40.2, 109.2\}$ | 90 | 0.01, 0 | 0.504 (0.004) | 0.615 (0.001) |
| Linear (n=11) | $f_0 \in \{2, 26.4, 50.8, 75.2, 99.6, 124.0, 148.3, 172.7, 197.1, 221.5, 245.9\}$ | 252 | 0, 0 | 0.592 (0.004) | 0.604 (0.001) |
| RNN (n=11) | - | 296 | 0, 0.05 | 0.617 (0.018) | 0.575 (0.002) |
| LSTM (n=6) | - | 298 | 0.001, 0 | 0.715 (0.019) | 0.546 (0.002) |

***Table 1: Characteristics and performance of different network architectures in two classification tasks.*** *RRNs consist of node frequencies $f_0$ spaced by factors of the golden ratio ("Golden", n=11 nodes), two (n=7 nodes), or Euler's number (n=5 nodes), or linear spacing from 2 Hz to 245.9 Hz (n=11 nodes). A standard recurrent neural network (RNN, n=11 nodes) and a long short-term memory (LSTM, n=6 nodes) network were also compared. Characteristics include the number of estimable parameters and values for two hyperparameters (weight decay and label smoothing coefficient). Performance includes accuracy classifying handwritten digits (MNIST) and spike train modulation (spike train).*

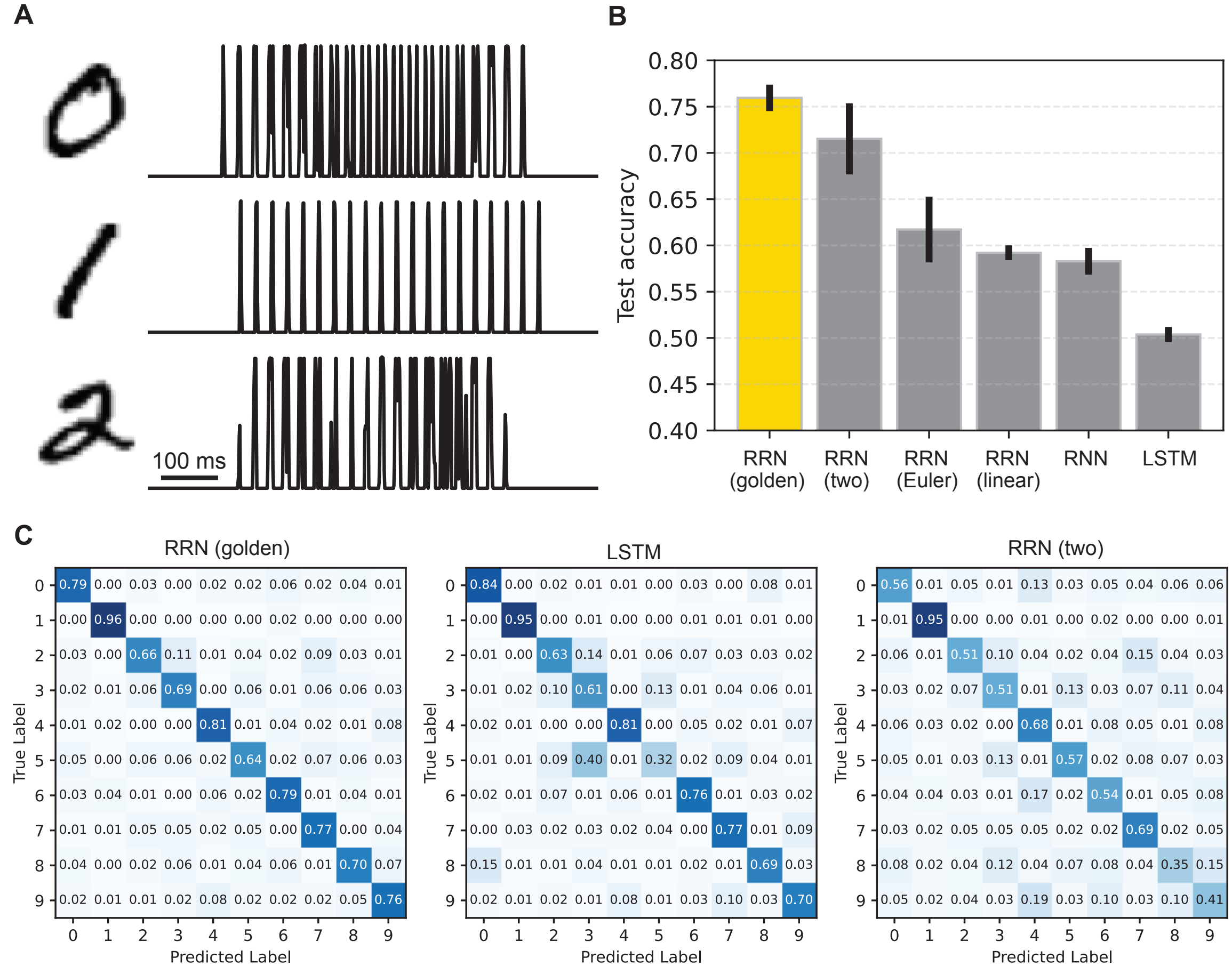

***Figure 2: The resonant recurrent network with golden ratio spacing classifies handwritten digits with highest accuracy. (A)*** *Example MNIST images and corresponding scanline time series.* ***(B)*** *Average test accuracies for each network architecture; black vertical lines indicate ±2 SEM.* ***(C)*** *Average confusion matrices for the golden RRN, LSTM, and RRN with factor of two spacing.*

**Examination of RRN dynamics provides insights into classification performance**

Inspection of the RRN outputs provides interpretable insights into classification performance. To illustrate this claim, we investigated how the golden RRN responded to different handwritten digits. To do so, we selected a trained golden RRN (uniform randomly from the 10 trained golden RRN architectures) and computed the response (the accumulated amplitude squared of each oscillator) to a randomly chosen digit from the test set (not used in training). For a fixed digit value (e.g., "1"), we repeated this process 100 times and determined the response mean and SEM across the 100 repetitions. Visual inspection of the example RRN output (the accumulated amplitude squared of each oscillator) revealed distinct responses to digits with high correct predictions (e.g., digits 1 and 4 in Figure 3). Digit 1 maintains a relative spectral peak at ≈36 Hz, corresponding to the nearly periodic scanline trace of this digit (see example in Figure 2A), with approximate period of 28 samples sampled at 1000 Hz (28 ms). Digits 2 and 3 have lower mean correct prediction ratios (0.66 and 0.69, respectively) and these two digits are often mislabeled (see off-diagonal elements in the confusion matrix of Figure 2C). The response to digit 2 (orange curve in Figure 3) overlaps with digit 3 (green curve in Figure 3). In this way, the RRN produces interpretable features, which provide insight into classification performance.

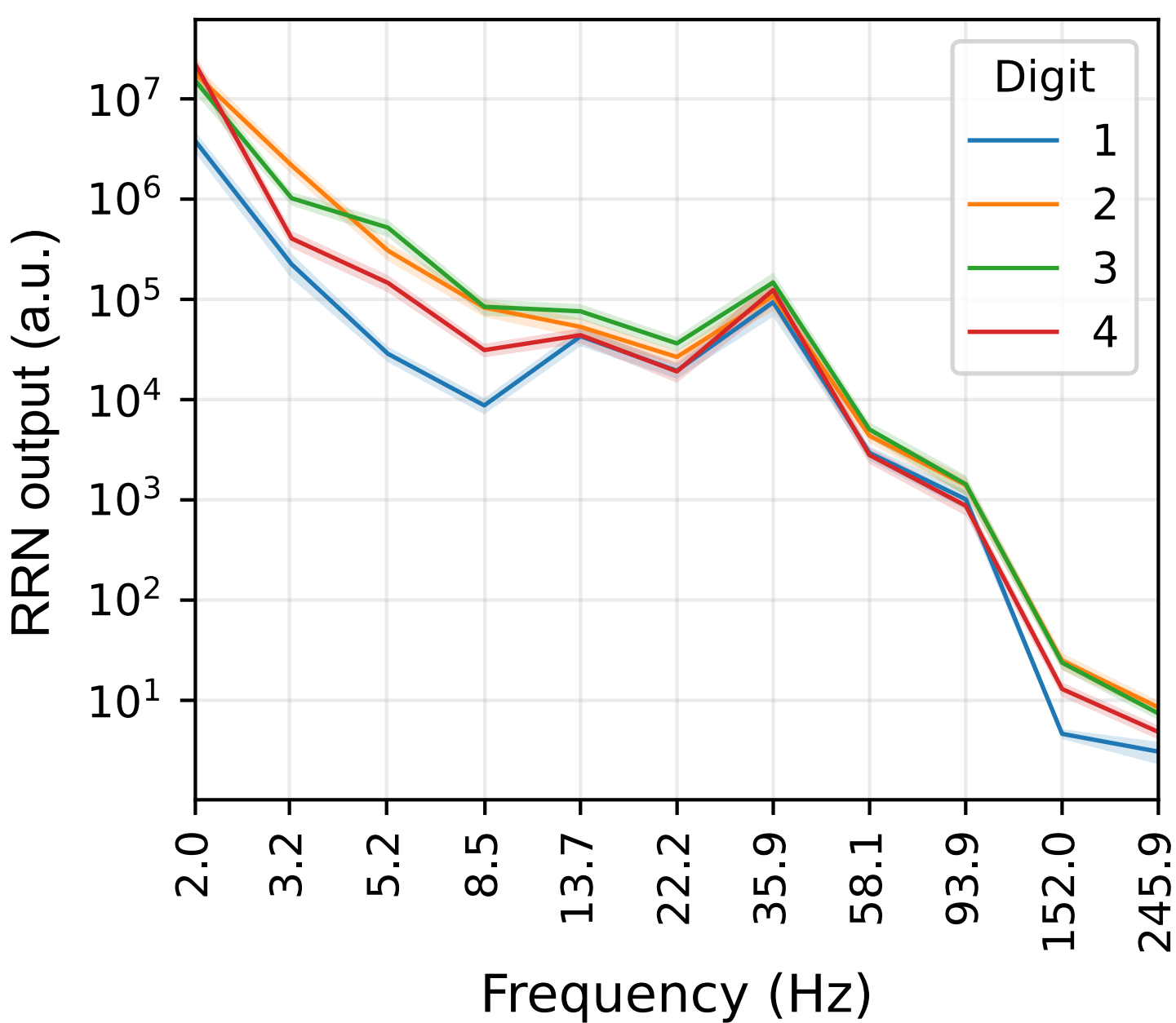

***Figure 3: RRN outputs depend on input digit and oscillator frequency.*** *The average (solid curves) and 95% confidence intervals (shaded regions) of the golden RRN output for four digits (see legend) from the testing dataset. The RRN response is frequency and digit dependent. Distinct responses (e.g., digits 1 and 4) are easier to classify than overlapping responses (e.g., digits 2 and 3).*

**Estimated recurrent connections indicate cross-frequency interactions**

Having found the RRN with golden ratio frequency spacing (i.e., the "golden RRN") performs best, we examined the resulting estimated connectivity between nodes. To do so, we trained the golden RRN with 100 realizations of the MNIST data, each using a different random seed. While the resulting connectivity matrix ($W$) varied across realizations (examples in Figure 4A), visual inspection of the mean connectivity matrix suggests more positive connection strengths from lower to higher frequency nodes (2-8.5 Hz to 22.2-93.9 Hz, solid black square in Figure 4B) compared to connection strengths from higher to lower frequency nodes (dashed black square in Figure 4B). Computing the average connectivity value for low-to-high frequency connections and low-to-high frequency connections, we find different weight distributions (Figure 4C). The mean low-to-high frequency connection strength exceeds the mean high-to-low frequency connection strength (t-test, $t = 12.1$, $p \approx 1e^{-25}$).

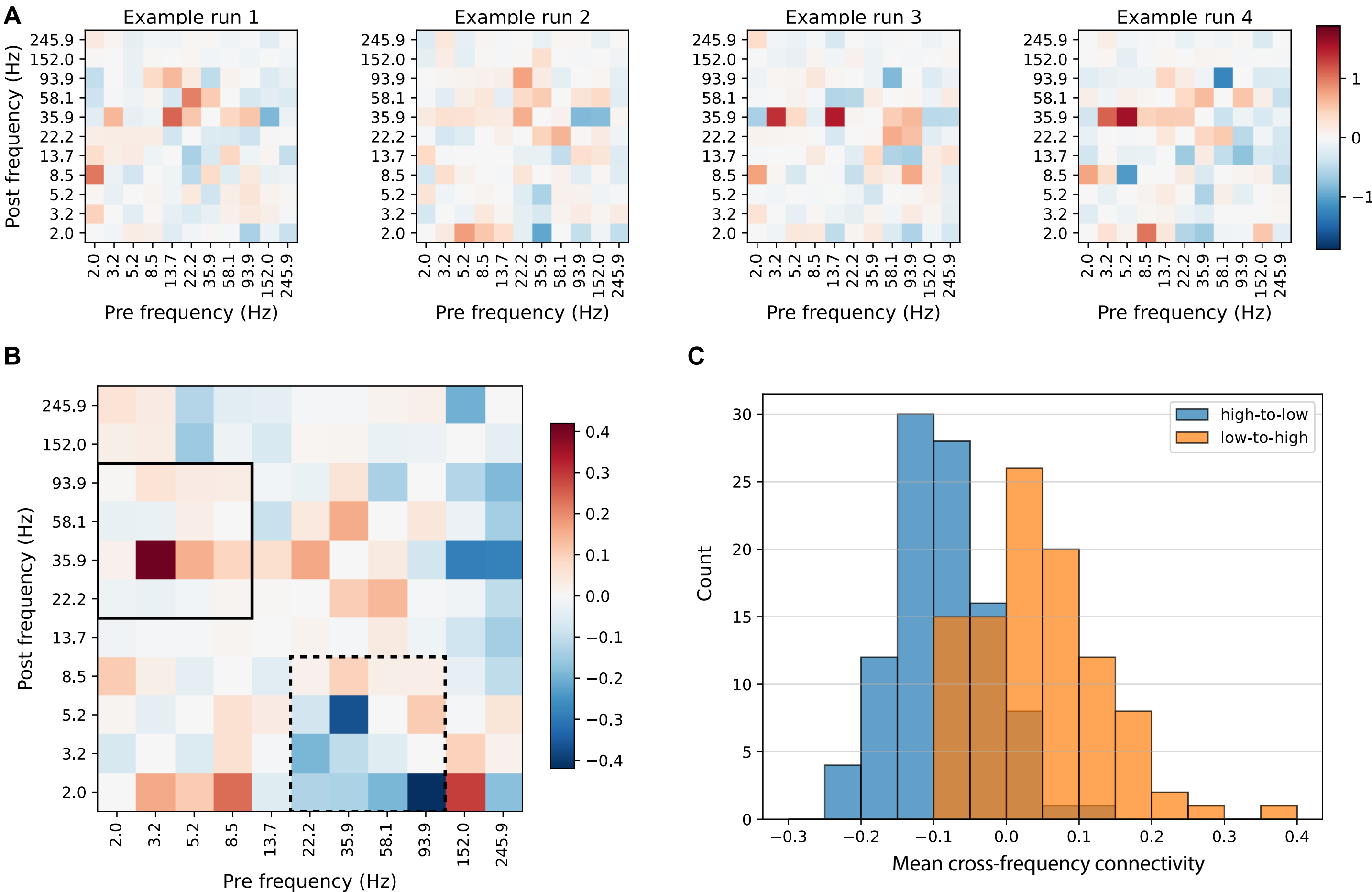

***Figure 4. Network connectives learned during training exhibit cross-frequency interactions. (A)*** *Example connectivity matrices $W$ estimated during different training runs.* ***(B)*** *Average connectivity matrix over all training runs. Black squares indicate low-to-high frequency connectivity (solid) and high-to-low frequency connectivity (dashed).* ***(C)*** *Distributions of average cross-frequency coupling strengths from low-to-high (orange) and high-to-low (blue).*

**Golden ratio spacing in the resonant recurrent network classifies different spike train patterns**

To illustrate RRN performance in a task more closely related to neuronal time series data, we considered the challenge of separating two different spike train patterns. The patterns consisted of brief (100 ms, sampling rate 1000 Hz) spike train data simulated in two conditions: (i) random spiking with an average firing rate of 100 Hz or (ii) random spiking with an average firing of 100 Hz modulated by a sinusoid with frequency chosen uniformly between 10 Hz and 50 Hz; see *Methods: Simulated spike train data*. Visual inspection of example traces demonstrates the difficultly of distinguishing between the constant and modulated scenarios (Figure 5A). To classify these spike train data, we trained and tested each network architecture using the fixed hyperparameters (weight decay and label smoothing coefficient) selected for the MNIST data (see *Methods: Hyperparameter selection*). Training estimated the connectivity between nodes ($W$ in Equation 3) and parameters for the linear classifier and batch normalization. To match comparisons across architectures, we trained (N=1000 examples), validated (N=200 examples), and tested (N=200 examples) each architecture using the same simulated spike train data. We repeated the entire training and testing process for 1000 realizations of the simulated spike train data. Consistent with the results for the MNIST dataset, we found that the RRN with frequencies separated by factors of the golden ratio - the golden RRN - performed with higher accuracy than the other architectures (Table 1, $p < 5e^{-11}$, Figure 5B). Confusion matrices demonstrated comparable or superior performance of the golden RRN compared to the other architectures (Figure 5C). Inspection of example RRN output (the accumulated amplitude squared of each oscillator) revealed an increased response of oscillators in the 10-50 Hz frequency range for the modulated data, as expected for these simulated spike trains (Figure 5D). Finally, the trained connectivity matrices ($W$, Figure 5E) demonstrated increased low-to-high frequency connection weights compared to high-to-low frequency connection weights (Figure 5F; t-test, $t = -4.9, p = 1.2e^{-6}$,), again consistent with the MNIST classification results. We conclude that that golden RRN performed consistently for the spike train data and the MNIST data. In both cases, the golden RRN outperformed the other architectures, produced interpretable features, and learned recurrent connectivity with the similar patterns of cross-frequency coupling.

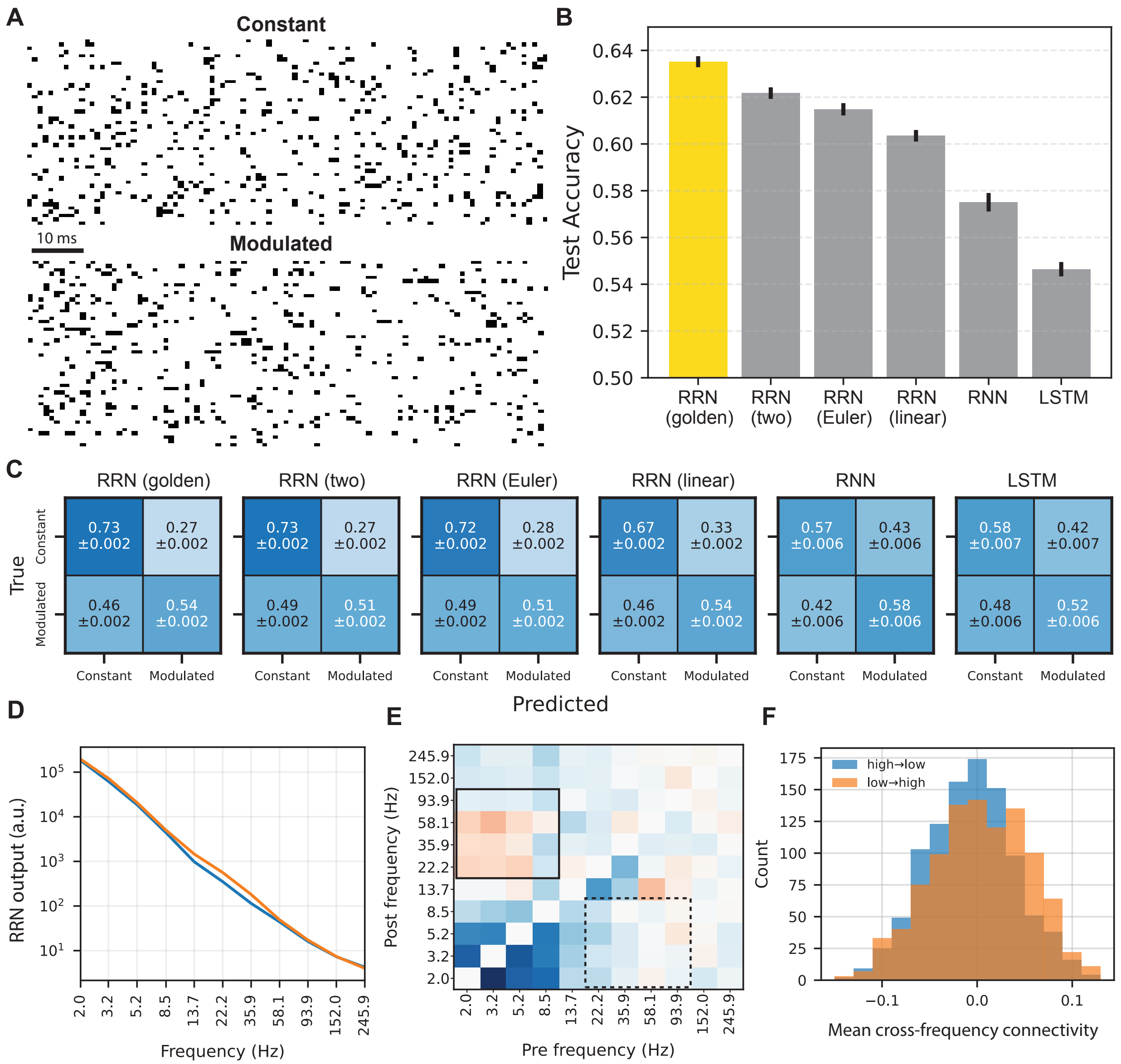

***Figure 5. The RRN with golden frequency spacing classifies spike train patterns. (A)*** *Example constant and modulated spike train patterns. Each row indicates one spike train realization (duration 100 ms). Fifty examples (rows) are shown.* ***(B)*** *Average test accuracy for each network architecture; black vertical lines indicate* $\pm 2$ *SEM.* ***(C)*** *Average confusion matrices (± SEM) for each network architecture.* ***(D)*** *Average (solid curves) and 95% confidence intervals (shaded regions) of the golden RRN output for each spike train condition (constant: blue, modulated: orange).* ***(E)*** *Average connectivity matrix over all training runs. Black squares indicate low-to-high frequency connectivity (solid) and high-to-low frequency connectivity (dashed).* ***(F)*** *Distributions of average cross-frequency coupling strengths from low-to-high (orange) and high-to-low (blue) frequencies.*

## Discussion

We propose a resonant recurrent network (RRN) consisting of damped oscillatory elements, which we express as a traditional recurrent neural network with an additional history dependent term. We show that the RRN outperforms standard recurrent neural network approaches to classifiy hand-written digits and simulated spike train data. Compared to existing recurrent neural network approaches, nodes in the RRN intrinsically oscillate and the system is simple to express (see Equation 3) and parameterize; the main parameter choice is the center frequency of each oscillator. Moreover, the structure and output of the RRN are directly interpretable; the RRN consists of damped, coupled oscillators and the outputs correspond to the summed amplitudes squared. Well-established constraints enforce stable node dynamics with fading (or short-term) memory and the input signals require no elaborate pre-processing. Finally, the RRN benefits from biological knowledge – the node frequencies can be chosen to follow the observed distributions of brain rhythms *in vivo*.

We propose two interpretations of the RRN: either as a network of discrete-time recurrent neural network elements with two history dependent terms, or as a network of continuous-time, coupled, damped, driven oscillators. Each interpretation offers insights into network performance and design. Typical recurrent neural networks implement first order discrete dynamical systems [43], [44], [45], [46], [47] consistent with autoregressive models of order one (AR(1)). The spectrum of an AR(1) consists of a low frequency peak and each node acts to low-pass filter input signals [37]. In this way, the node dynamics in typical recurrent neural networks are limited and cannot resonate with narrowband higher frequency oscillations. The two history dependent terms in the RRN extend these typical systems to an autoregressive model of order two (AR(2)), thus extending the capabilities of each node to narrowband filtering. Alternatively, conceptualizing the RRN as a network of damped oscillators provides physical intuition and offers opportunities to connect the neural network dynamics to existing well-developed theories of coupled oscillators [48], [49], [50], [51]. Application of these existing theories may facilitate a deeper understanding of the RRN dynamics, providing insights difficult to obtain in typical deep neural network formulations.

We showed that frequencies separated by factors of the golden ratio outperformed alternative proposed frequency spacing scenarios (e.g., logarithmic [41] or factor of two [42]) in the RRN. These results are consistent with existing theoretical [32], [52], [53], [54] and experimental [39], [55] works, which suggest that golden ratio spacing supports both independent communication channels and coupled cross-frequency interactions. Training the node-to-node connectivity matrices using a standard artificial neural network approach (backpropagation through time), we find evidence of cross-frequency interactions, with low-to-high frequency oscillators more positively coupled than high-to-low frequency oscillators. These cross-frequency interactions may be especially important for characterizing the sharp, abrupt changes characteristic of the scanline MNIST or simulated spike train time series data. How these interactions between simulated damped oscillators relate to the biological mechanisms of interacting neurons or neural populations remains a topic for future investigation.

The proposed RRN differs from existing oscillatory recurrent neural networks [21], [22] in the following ways. First, the RRN formulation provides a direct link from the physical system – damped oscillators – to a simple extension of existing recurrent neural network models. The fundamental insight is to augment the traditional recurrent neural network model ($x_t = \alpha\, x_{t-1}$) to include a second order term ($x_t = \alpha_1\, x_{t-1} + \alpha_2\, x_{t-2}$). Second, we choose oscillator frequencies in the RRN to match the spacing between the brain's frequency bands observed *in vivo* [32], [41], [42]. Doing so, we find that golden ratio spacing outperforms alternative scenarios and standard recurrent neural networks (LSTM and Elman RNN) with a comparable number of parameters. Third, the number of estimated parameters in the RRN is small; determining the recurrent connections between $N$ nodes requires estimation of $N^2 - N$ parameters (no self-connections), with $N \leq 11$. Fourth, the RRN produces interpretable outputs, the accumulated amplitude squared (or response) of each oscillator and investigation of these responses provides insight into classification performance.

Here we used training data to estimate connectivity between RRN nodes. An alternative approach is to constrain or impose network connectivity motivated by physical or biological insight. For example, replacing the inferred network architecture with a finite difference approximation of the Laplacian, the activation function with the node activity, and limiting the frequency differences between nodes (e.g., $\Delta f \approx 0$) would simulate diffusive coupling between approximately homogenous damped, driven oscillators. In this way, the *in silico* RRN connectivity and dynamics may be altered to approximate existing physical reservoir computing strategies (e.g., liquid media [56], [57], memristors [58], nanoscale magnetic oscillators [59], or human brain organoids [60]) and, for example, support waves of neural activity [61], which may implement computational principles similar to the transformer networks of large language models [62]. Updating the network architecture to use *in vivo* brain connectivity [63], [64] would incorporate additional biological features, and the nonlocal, distributed interactions may better support computations than the local interactions of simpler physical systems [56].

We note that an appropriate choice of network architecture could combine two nodes, each with one history dependent term (i.e., $t - 1$), to implement second order dynamics. This connectivity could be imposed or, perhaps, result from training. However, in this case, interpretation of the node activity becomes less direct. For example, we might interpret some nodes as representing the position and other nodes the velocity of a dynamical system. While mathematically equivalent, this first order system is conceptually more complex. Here, we instead impose second order dynamics at each node, or equivalently dynamics that depend on two history dependent terms at $t-1$ and $t-2$. An advantage of this framework is a direct physical interpretation and connection to a clear physical mechanism (i.e., a damped oscillator) or to narrowband filtering at each node.

Traditional artificial neural networks draw inspiration from the brain's neuronal dynamics. Here, we continue this approach by mimicking an additional, well-established feature of neuronal dynamics: rhythms [4]. Most artificial recurrent neural networks implement first order dynamics and therefore do not intrinsically oscillate. The brain, however, oscillates in both health and disease [23], [65], [66]. Why biological neurons and brain circuits oscillate remains unknown. Oscillations have been proposed to support brain processes including temporal filtering via resonance [14], [15], communication through coherence [16], [17], and cross frequency interactions [18], [19]. More generally, biological brains exist in an oscillatory world and therefore may utilize oscillatory components to exploit these oscillatory inputs. Because many arbitrary signals can be represented as a summation of sinusoids, developing a biological system that decomposes signals as sinusoids supports a rich feature space. Alternatively, biological neural networks may have no other choice as the only available components (biological neurons) necessarily oscillate [67], [68]. In that case, artificial brains, which are not restricted to an oscillatory environment or oscillatory components, may have an advantage over biological brains. However, recent work directly comparing oscillatory and non-oscillatory neural networks suggests that oscillations may support performance [22]. Beyond oscillations, our biological neural networks may possess intrinsic advantages not immediately accessible to artificial neural networks. If so, protecting these advantages may serve strategic outcomes if biological and artificial brains must ultimately compete [69].

## Methods

### *MNIST Dataset*

The MNIST dataset [70] consisted of ten handwritten digits from 0 to 9. The training set contained 60,000 images and the testing set 10,000 images (Table 1). Each digit consisted of a 28 x 28 matrix of intensity values, which we transformed (scanline order) to a (1 x 784) time series with assumed sampling frequency of 1000 Hz and rescaled to values between 0 and 1 (Figure 2A).

| *Digit* | *Training Count* | *Testing Count* |
|---|---|---|
| *0* | 5923 | 980 |
| *1* | 6742 | 1135 |
| *2* | 5958 | 1032 |
| *3* | 6131 | 1010 |
| *4* | 5842 | 982 |
| *5* | 5421 | 892 |
| *6* | 5918 | 958 |
| *7* | 6265 | 1028 |
| *8* | 5851 | 974 |
| *9* | 5949 | 1009 |
| *Total* | 60000 | 10000 |

### *Model training*

All models were trained using a common PyTorch [71] pipeline with supervised minibatch optimization. For the MNIST data, for a chosen random seed, a fixed train/validation split was generated by shuffling the 60,000 training examples and holding out 5,000 for validation; the standard 10,000-example MNIST test set was used only for final evaluation. For the synthetic spike train data, for a chosen random seed, synthetic data were generated with constant or modulated firing rate, with 1000 training examples (500 per class), 200 validation examples (100 per class), and 200 testing examples (100 per class). Models were trained for 30 epochs (batch size 128) with AdamW, cross-entropy (optional label smoothing), global gradient clipping (norm 1.0), and OneCycle LR scheduling (cosine annealing, stepped each minibatch). After each epoch, validation accuracy was computed; the best-validation checkpoint was saved, reloaded after training, and evaluated on test accuracy. For the RRN configurations, the diagonal of the recurrent weight matrix $W$ was constrained to zero (preventing self-connections) by applying a post-update projection after each optimizer step.

*Hyperparameter selection*

We selected hyperparameters via a grid search performed separately for each architecture variant (each RRN type, LSTM, and the recurrent neural network). For each candidate setting, models were trained for 30 epochs with the same training loop and objective (multiclass cross-entropy with optional label smoothing), varying only the optimizer/loss hyperparameters: learning rate $\in \{3 \times 10^{-4}, 10^{-3}, 3 \times 10^{-3}\}$, weight decay $\in \{0, 10^{-3}, 10^{-2}\}$, and label-smoothing coefficient $\in \{0, 0.05\}$. Each grid point was evaluated across three independent random seeds; for each seed, the model was re-initialized, and 5000 MNIST examples were held out for validation. During training, validation accuracy was computed for each epoch and the checkpoint achieving the maximum validation accuracy over the full training trajectory was retained. After completing the grid, hyperparameters were selected per architecture by maximizing the mean best validation accuracy across the three seeds, requiring that all three seeds were present for a candidate configuration. The test set was not used for hyperparameter selection; the chosen hyperparameters were subsequently fixed for subsequent evaluations. We note that, for all network architectures, a learning rate of $3 \times 10^{-3}$ optimized performance.

*Simulated spike train data*

Spike train data were simulated as realizations of a Poisson process with either constant intensity of 100 Hz or modulated intensity. For the modulated intensity,

$$\lambda_t = f_c + f_c \ \sin\left(2\,\pi\, f_m\, t\right)$$

where $f_c = 100$ Hz and $f_m$ was selected (uniformly) between 10 Hz and 50 Hz for each spike train realization.

*Statistical tests*

To test for differences in accuracy between models, we estimated linear models with output test accuracy and categorical predictors corresponding to each model type (reference category set to the golden RRN) and a constant predictor.

*Code availability*

All methods required to reproduce the results in this paper are available at https://github.com/Mark-Kramer/Resonant-Reservoir-Network for reuse and further development.